\documentclass[aps,prb,preprint,amssymb,superscriptaddress,showpacs]{revtex4-1}

\usepackage{verbatim}
\usepackage[english]{babel}
\usepackage{amsmath,amsthm}
\usepackage{graphicx}
\usepackage{subfigure}
\usepackage{epsfig}
\usepackage{threeparttablex}
\usepackage{natbib}
\begin{document}
\title{Exchange--correlation effects in the monoclinic to tetragonal phase stabilization
       of Yttrium--doped $ZrO_2$: a first--principles approach}

\author{Davide Sangalli}
\affiliation{MDM Lab - IMM - CNR via C. Olivetti, 2 I-20864 Agrate Brianza (MB) Italy, European Union}
\affiliation{Consorzio Nazionale Interuniversitario per le Scienze dei Materiali (CNISM)} 

\author{Alberto Debernardi}
\affiliation{MDM Lab - IMM - CNR via C. Olivetti, 2 I-20864 Agrate Brianza (MB) Italy, European Union}

\date{\today}
\begin{abstract}

We describe, within an ab--initio approach, 
the stabilization of the tetragonal phase vs. the monoclinic one
in Yttrium--doped Zirconia. The process is believed
to be influenced from different mechanisms.
Indeed we show that there is a delicate balance between the change in
electrostatic and kinetic energy and exchange--correlation effects. In the 
tetragonal phase the perturbation induced by doping is better screened at 
the price of sacrificing correlation energy.
Our work opens the opportunity to use the
same approach to predict the tetragonal phase stabilization of materials
like Zirconia or Hafnia, with different and less characterized dopants.

\end{abstract}

\pacs{64.70.K-, 71.15.Mb, 81.05.Je, 81.30.Bx} 

\maketitle

\section{Introduction}
Zirconia ($ZrO_2$) is a hard and usually colorless material with a wide range of
technological applications\cite{Heurer1981-84}.
Being corrosion resistant it is used as a dental material and because
of its low cost, durability, and close visual
likeness to diamond, it is widely used to synthesize artificial gems. 
It is used as a thermal barrier in coating engines due to its high resistance.
The addition of cations (as for example $Y^{3+}$) induces the generation of oxygen vacancies
for charge compensation which makes it useful as oxygen sensor. Moreover
Zirconia is an high dielectric constant (high-$\kappa$) material with potential applications
in the micro--electronics. Finally, very recently, transition--metals doped Zirconia has been
predicted to be a dilute magnetic semiconductor (DMS) with high Curie Temperature $T_c$
with potential applications in the field of spintronics~\cite{Ostanin2007}; while pure and doped
$ZrO_2$ has been proposed as a candidate material for resistive switching memories
devices~\cite{Wu2007,Zhang2010} (ReRAM) exploiting the high vacancies mobility of the system.

Pure $ZrO_2$ exhibits at ambient pressure three polymorphisms. The monoclinic (M) phase is stable at low temperature
and is the less symmetric structure with the $Zr^{4+}$ ions exhibiting sevenfold coordination. Between
1440 and 2640 K the tetragonal (T) phase, with eightfold coordinated $Zr^{4+}$ ions is stable.
Finally above 2640 K, till melting temperature ($\approx 3000 $K), the most symmetric cubic (C) phase
is stabilized\cite{Rignanese2005,Dewhurst1998}. The only difference between the (C) and the (T) phase is
a distortion of the oxygen sub--lattice with a spontaneous symmetry breaking. The (T) and (C) phases are more used in
technological applications and both the (T)/(C) structures and the (T) $\rightarrow$ (C) doping induced
phase transition (DIPT) have been well characterized in the literature
both form the theoretical and the experimental point of
view~\cite{Stefanovich1994,Stapper1999,Ostanin2002A,Ostanin2002B,Lau2009,Goff1999,Aldebert1985,Howard1988,Lamas2000,Clarke2003,Rignanese2005},
with $Y^{3+}$ ions the most used and studied dopants.

On the other hand the (M) $\rightarrow$ (T) DIPT is much less characterized, especially from the theoretical point
of view. It presents a volume change of about $3-4 \%$ that causes extensive cracking in the material.
This behavior destroys
the mechanical properties of fabricated components and makes pure Zirconia useless
for structural or mechanical application. Moreover the (T) phase,
is metastable in pure and lightly doped $ZrO_2$
over a very long time; a growth sample of doped Zirconia must be annealed in order to check if 
the reached tetragonal phase is stable or meta--stable. Hence a better understanding of 
the (M) $\rightarrow$ (T) DIPT is desirable. In this paper we address the problem
from a first principle perspective and, in particular, we
show that the DIPT is a balance of the mean field (MF) description
(i.e. kinetic plus electrostatic: $H=T+V^{ext}+V_H[\rho]$) with 
exchange--correlation effects which cannot be captured by simplified models.

\section{First principles description.}
\subsection{Computational details.}

\begin{table}
\linespread{1.}
\parbox{0.6\textwidth}{\caption{Theoretical and experimental cell parameters of $ZrO_2$ in the monoclinic and in the tetragonal phase.}}
\label{tab:ZrO2_parameters1}
\begin{tabular*}{0.6\textwidth}{@{\extracolsep{\fill}} l c | c c | c c }
\hline
\hline
\multicolumn{2}{c}{Present work} &  \multicolumn{2}{c}{Theory} & \multicolumn{2}{c}{Experiment}    \\
\hline
\multicolumn{6}{c}{(M) $ZrO_2$}\\
         &                       &  Ref.[\onlinecite{Jiang2010}]
                                                 & Ref.[\onlinecite{Stapper1999}]
                                                               & Ref.[\onlinecite{Howard1988}]
                                                                                 &  Ref.[\onlinecite{Wyckoff1966}]\\
a  (\AA) &  5.18                 &  5.05         &             & 5.15            & 5.15            \\
b/a      &  1.011                &  1.027        &             & 1.012           & 1.012           \\
c/a      &  1.037                &  1.028        &             & 1.023           & 1.032           \\
$\beta$  &  $99^\circ 10'$       & $99^\circ 5'$ &             & $99^\circ 14'$  & $99^\circ 14'$  \\
\hline
\multicolumn{6}{c}{(T) $ZrO_2$}  \\
a (\AA)  &  5.11                 & 5.02           & 5.03       &                 & 5.07            \\
c/a      &  1.030                & 1.014          & 1.017      & 1.026           & 1.018           \\
\hline
\hline
\end{tabular*}
\end{table}

We work within density functional theory~\cite{Hohenberg1964,Kohn1965}
(DFT) in the generalized gradient approximation~\cite{PBE} (GGA)
with ultrasoft pseudopotentials~\cite{Vanderbilt1990,rrkjus1990}.
Both for Yttrium an Zirconium the pseudopotentials includes semicore
electrons and non--local core correction.
The Quantum Espresso~\cite{QE} package
is used to solve the Kohn--Sham~\cite{Kohn1965} (KS) equations for a super--cell with 96 atoms (down to 92 when oxygen 
vacancies are considered). We checked that a cut--off of 35 Ry for the wave--functions
and 400 Ry for the augmentation charge, a k-point grid $2\times 2\times 2$ are needed in order to have
the energy difference between the tetragonal and the monoclinic phase converged to up to $5.0*10^{-4}$ eV per $ZrO_2$
molecular unit (m.u.), where the energy difference is of the order of $0.1$ - $0.01$ eV/m.u.\ .

As starting point we relaxed atomic positions and structure of pure $ZrO_2$ with 12 atoms in the unit
cell of the (T) and (M) phases. The (C) phase is unstable and is not considered at zero
doping~\cite{NoteD}. The results, reported in Tables~\ref{tab:ZrO2_parameters1}-\ref{tab:ZrO2_parameters2}, are in 
good agreement with experimental data and previous theoretical works within the DFT error ($\approx1-2$\%).

\begin{table}
\linespread{1.}
\parbox{0.7\textwidth}{\caption{Theoretical and experimental internal structural parameters of $ZrO_2$ in
                                the Wyckoff~\cite{Wyckoff1966} notation. For the tetragonal structure
                                the only free parameter is the $z$ coordinate of the oxygen atoms
                                $(0.0,\ 0.5,\ 0.25-d_z)$}}
\label{tab:ZrO2_parameters2}
\begin{tabular*}{0.7\textwidth}{@{\extracolsep{\fill}} l c | c | c c }
\hline
\hline
         &    Present work       &  Theory                          &  \multicolumn{2}{c}{Experiment}          \\
\hline
\multicolumn{5}{c}{(M) $ZrO_2$}                             \\ 
         &                       &  Ref.[\onlinecite{Stapper1999}]  &  \multicolumn{2}{c}{Ref.[\onlinecite{Howard1988}]} \\
$Zr$     &  (0.276,0.044,0.210)  &  (0.277,0.043,0.210)             &  \multicolumn{2}{c}{(0.275,0.040,0.208)} \\
$O_I$    &  (0.065,0.327,0.350)  &  (0.064,0.324,0.352)             &  \multicolumn{2}{c}{(0.070,0.332,0.345)} \\
$O_{II}$ &  (0.451,0.757,0.475)  &  (0.450,0.756,0.479)             &  \multicolumn{2}{c}{(0.450,0.757,0.479)} \\
\hline
\multicolumn{5}{c}{(T) $ZrO_2$}  \\
         &                       &                                  &  Ref.[\onlinecite{Aldebert1985}]
                                                                                             & Ref.[\onlinecite{Fukuhara1993}]  \\
$d_z$    &  0.057                &  0.044                           &  0.057                 & 0.042     \\
\hline
\hline
\end{tabular*}
\end{table}

We correctly find that the (M) phase is favored with an energy difference 
$\Delta E_{M-T}=0.109$ [eV/mol] in agreement with previous works
(0.063 [eV/mol]\cite{Stapper1999}, 0.144 [eV/mol]\cite{Dewhurst1998}); the experimental estimation
is 0.063 [eV/mol]\cite{Achermann1975}.
		
\subsection{Yttria--stabilized Zirconia}

The (T) and the (C) phase of Zirconia can be stabilized with Yttria ($Y_2O_3$) doping, with a phase
diagram for Yttria--stabilized Zirconia (YSZ) which has been extensively characterized experimentally.
Also the (T) and (C) phases of Y--doped Zirconia have been studied within a theoretical
approach in a number of works~\cite{Stapper1999,Ostanin2002A,Ostanin2002B},
while no systematic study of the Y--doped (M) phase or of
the (M) $\rightarrow$ (T) DIPT has been reported in the literature.

\begin{figure}[t]
 \begin{center}
 \subfigure[DOS $ZrO_2$ with $Y$ doping]{\includegraphics[width=0.475\textwidth]{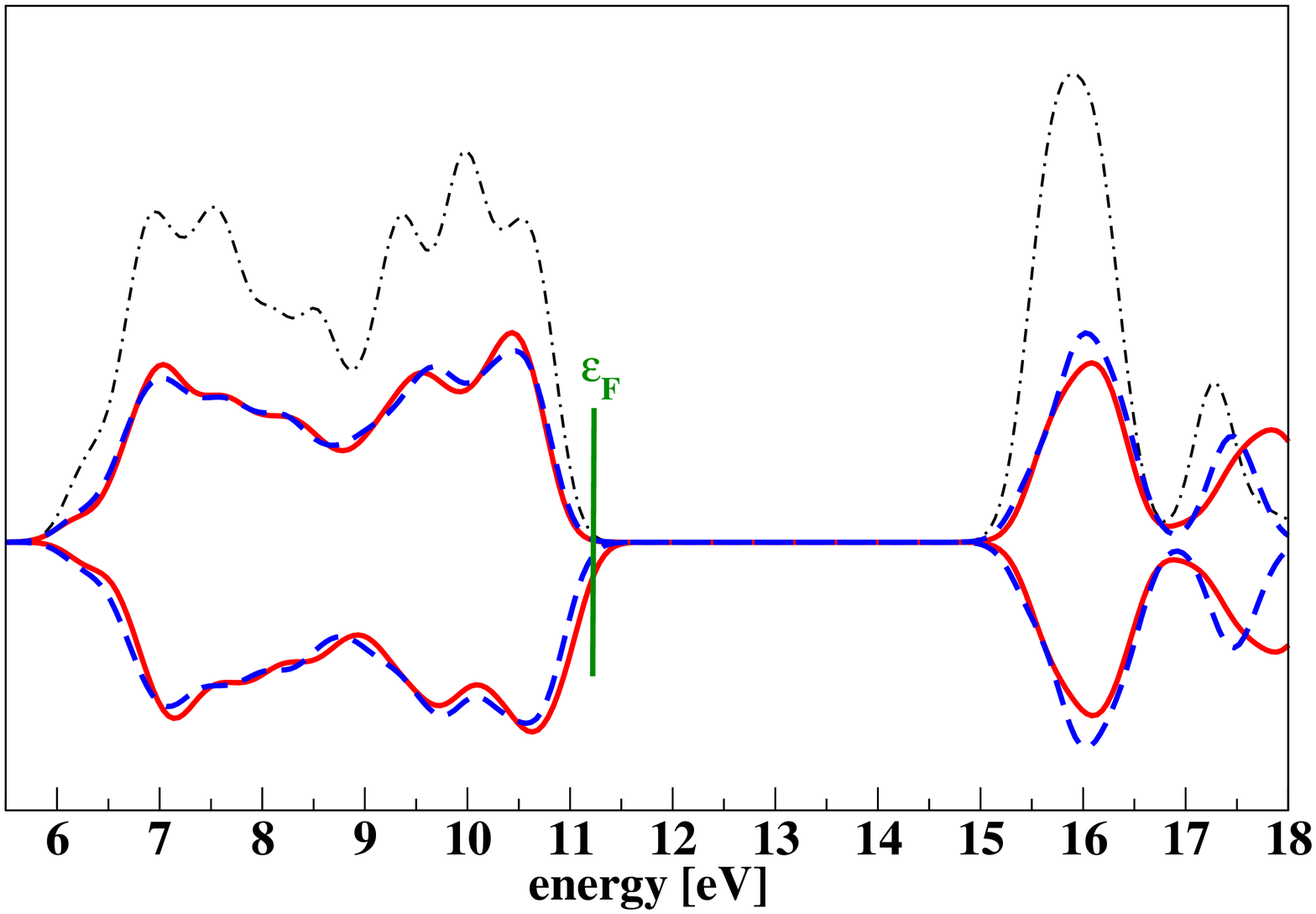}\label{fig:dos_ZrO2+Y}}
 \subfigure[DOS $ZrO_2$ with vacancy]{\includegraphics[width=0.475\textwidth]{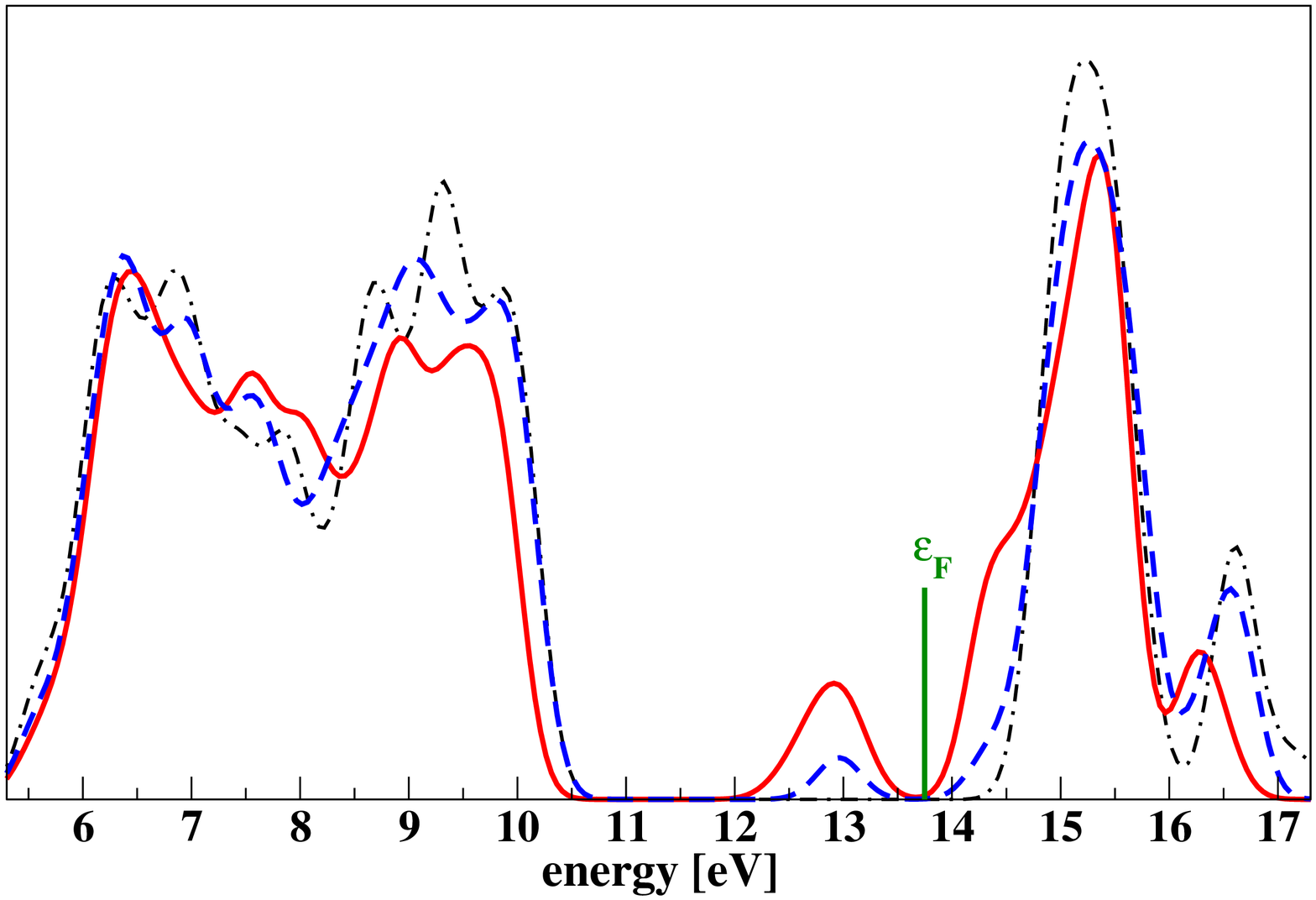}\label{fig:dos_ZrO2+vac}}
 \subfigure[DOS $ZrO_2$ with $Y_2O_3$ doping]{\includegraphics[width=0.475\textwidth]{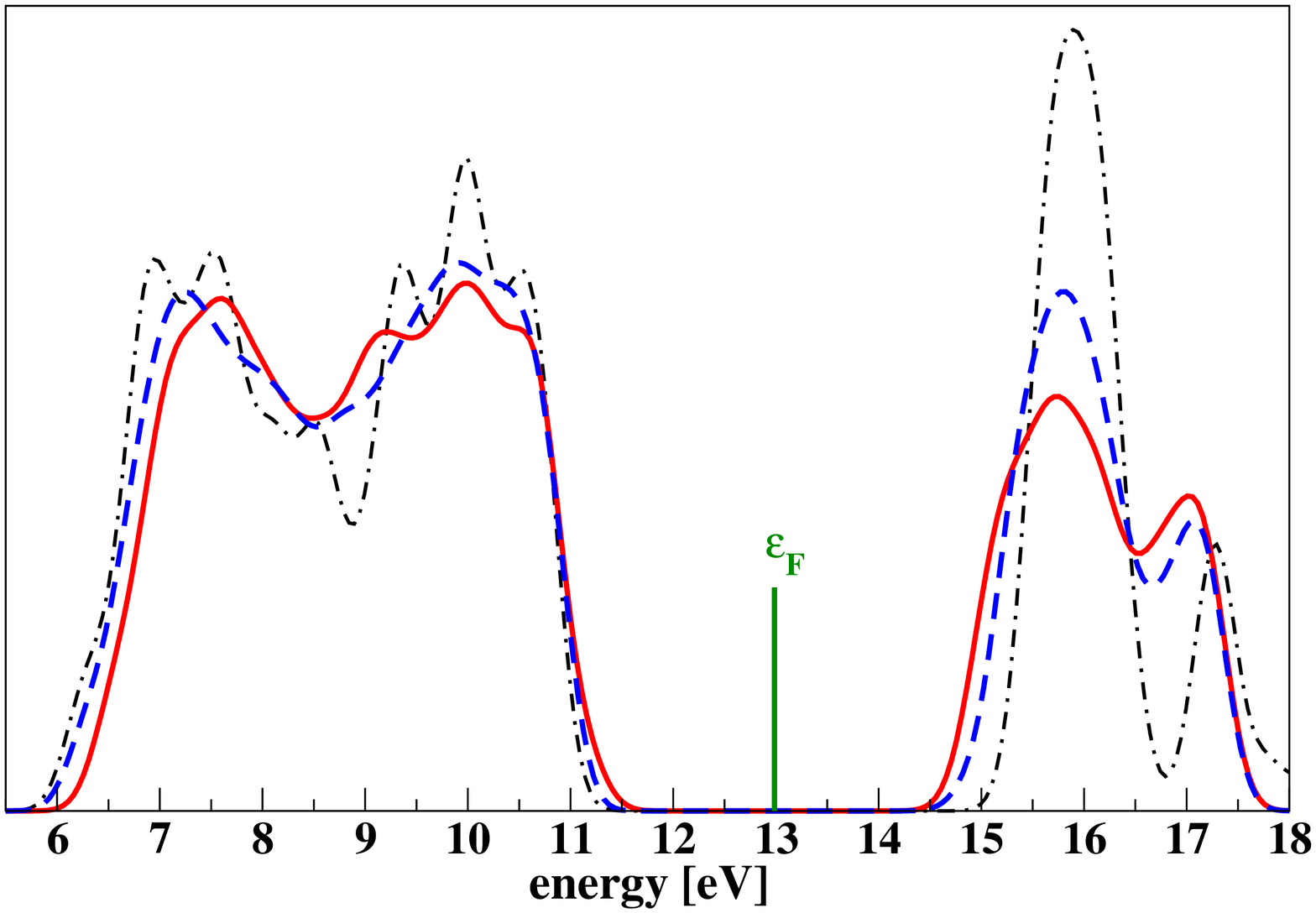}\label{fig:dos_ZrO2+Y2O3}}
 \subfigure[Energy gain per vacancy]{\includegraphics[width=0.475\textwidth]{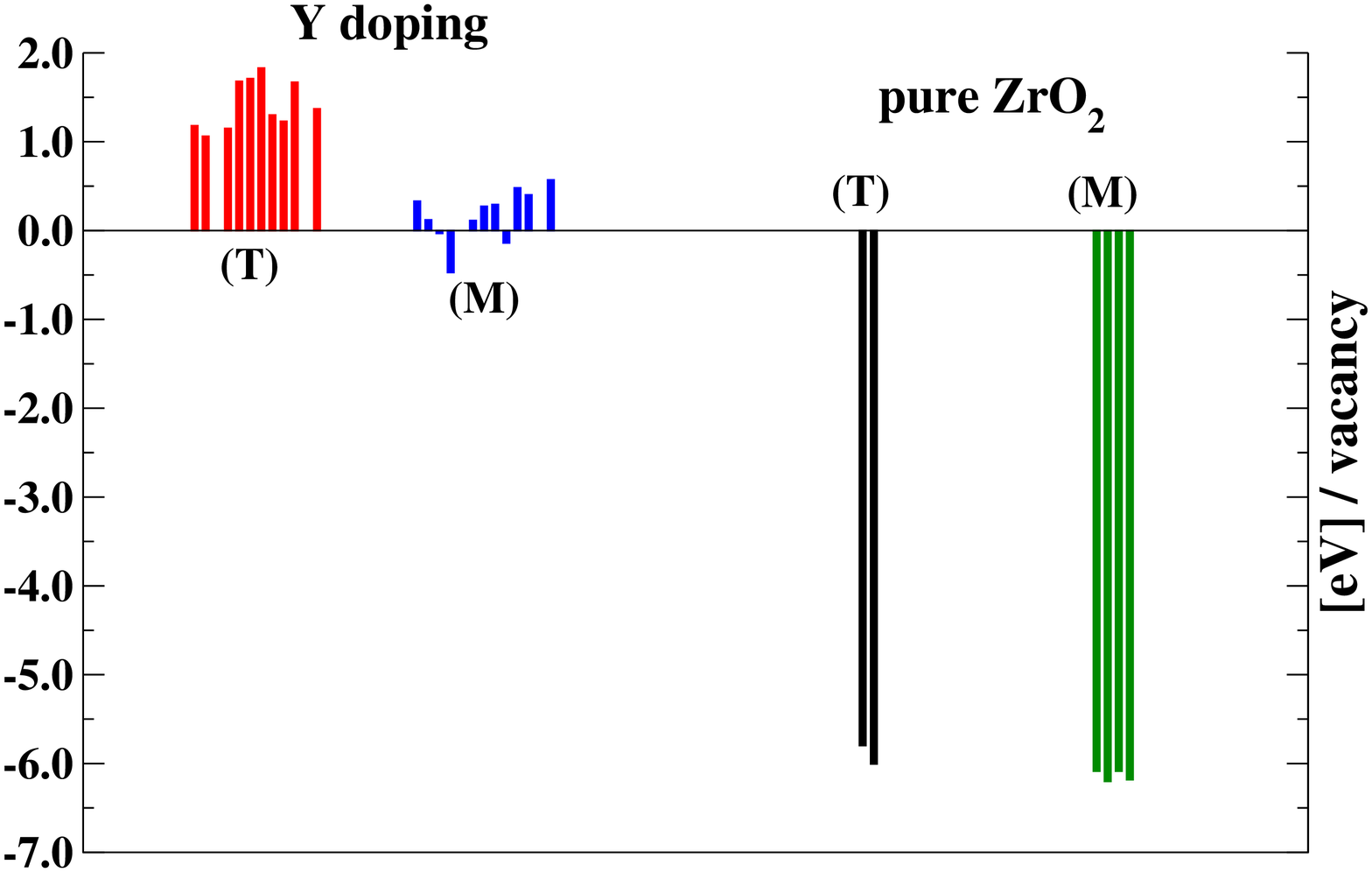}\label{fig:vac_energy}}
 \end{center}
  \linespread{1.}
 \caption{(colors online) Kohn--Sham density of states for $(a)$ Yttrium doped zirconia
          ($Zr_{1-x}O_2Y_x$), $(b)$ zirconia with oxygen vacancies ($ZrO_{2-y}V_y$) and
          $(c)$ zirconia with one vacancy each two Yttrium atoms ($Zr_{1-x}O_{2-y}Y_xV_y$) in the tetragonal phase.
          The DOS are at $x=0.25$ and $y=0.125$ (red continuous line) and $x=0.125$ and $y=0.0625$ (orange dashed line);
          the DOS of pure $ZrO_2$ (thinner black dot--dashed line) is also shown.
          The system doped with Yttrium gains energy when vacancies are created, releasing oxygen molecules,
          (panel $(d)$). Here $E_{gain}=1/y\{E(Zr_{1-x}O_2Y_x)-[E(Zr_{1-x}O_{2-y}Y_xV_{y})+y/2\mu(O_2)]\}$, with $x=2y$
          for the ``Y--doping'' case (on the left) and $x=0$ for pure $ZrO_2$ (on the right). The creation of
          vacancies is instead highly unfavored in the undoped system.}
 \label{fig:ZrO2+Y_and_Vac}
\end{figure}

$Y$--doping is known to induce oxygen vacancies. We show in Fig.~\ref{fig:ZrO2+Y_and_Vac} how the Kohn--Sham (KS)
density of states changes if vacancies are present both in pure and doped $ZrO_2$ crystals at
low doping~\cite{NoteC}.

Moreover we compute the energy the system gains
producing vacancies (see the caption of Fig.~\ref{fig:ZrO2+Y_and_Vac} for more details) for different doping concentrations. 
A correct modelling of the material must
consider the possible relative positions of the dopants (as substitutional defects) and the vacancies.
We found out that the relative position of Y atoms among themselves plays a minor role, in agreement
with the other works~\cite{Ostanin2002A}, and in our system we kept these
as far as possible to mimic uniform doping. Instead the position of vacancies respect to Y atoms or to other vacancies
influences significantly the total energy of the system.
Stapper et al.\cite{Stapper1999} reported that in the cubic system isolated vacancies tend to be next nearest neighbor (NNN)
to $Y$ atoms with an energy gain of $\Delta E_c\approx 0.34$ eV per vacancy against the nearest neighbor (NN) configuration in a supercell with
1 $Y$ atom and 1 vacancy. This result is confirmed also by Ostanin et al.\cite{Ostanin2002A} in a supercell with 2 $Y$ atoms and 1 vacancy
with $\Delta E_c\approx 0.3$ eV between the two configurations $(O$-$Y_1,O$-$Y_2)=(NN,NNNN)$ and $(O$-$Y_1,O$-$Y_2)=(NNN,NNN)$ in favor
of the second in the (C) phase and $\Delta E_t\approx 0.2$ eV in the (T) phase. Also in the present work we found $\Delta E_t\approx 0.12$ eV
considering 1 vacancy and 2 $Y$ atoms.

In the (M) phase two nonequivalent oxygen atoms exist, one with coordination $~3$ and another with coordination $~4$, and, accordingly
there are four possible configurations. we found that the case
$(NN,NNNN)x3$ favored of 0.21, 0.36 and 0.80 eV against $(NNN,NNN)x3$, $(NNN,NNN)x4$ and $(NN,NNNN)x4$ respectively; here the labels $x3$ and $x4$
refer to the coordination that would have the vacant oxygen. 
At the best of our knowledge, no results have been reported in the literature about the position of vacancies in the 
(M) phase.

\begin{figure}[t]
 \begin{center}
 \subfigure[\ Tetragonal]{\includegraphics[width=0.488\textwidth]{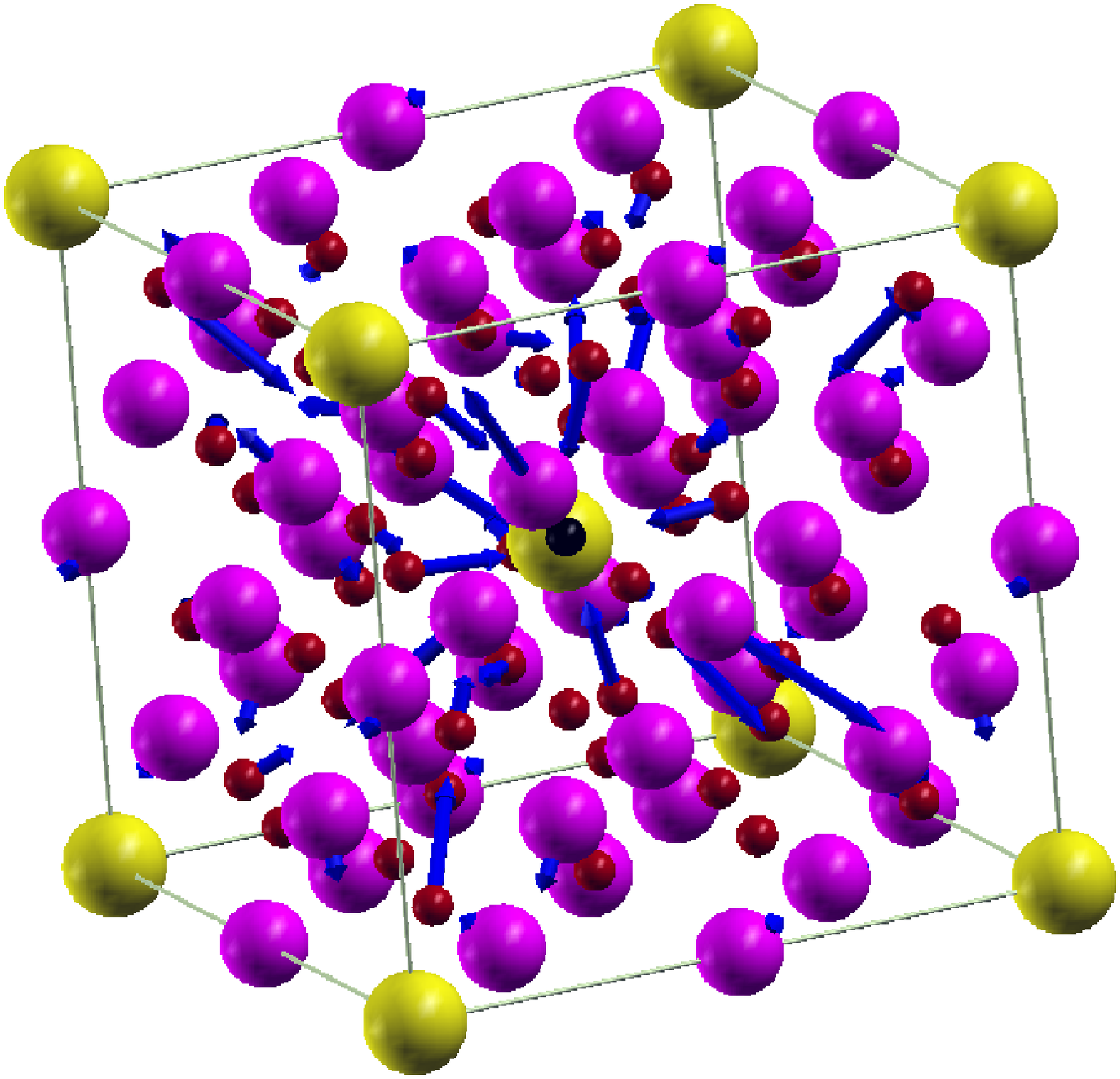}\label{fig:tetra}}
 \subfigure[\ Monoclinic]{\includegraphics[width=0.48\textwidth]{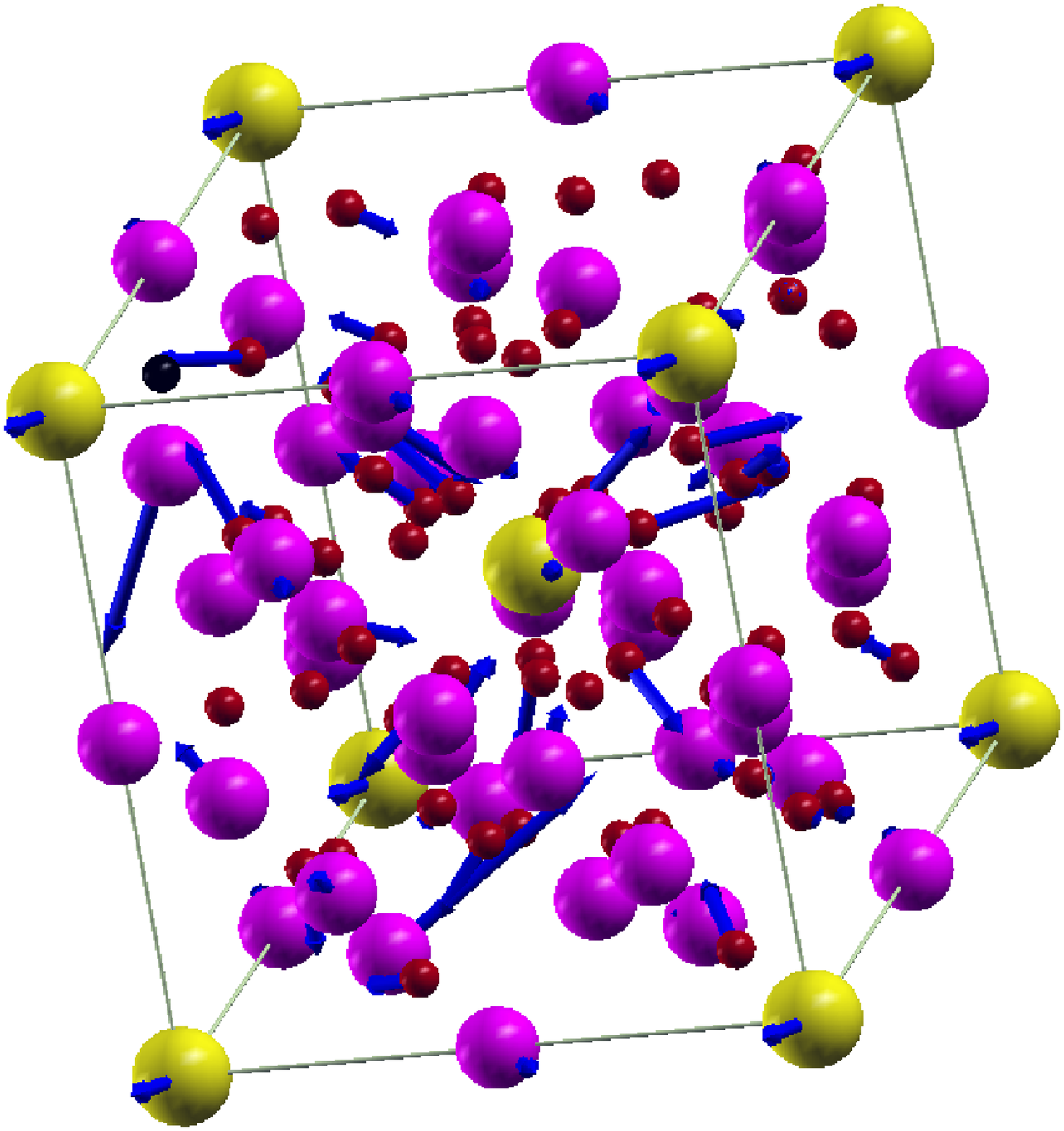}\label{fig:mono}}
 \end{center}
  \linespread{1.}
 \caption{(colors online) Tetragonal $(a)$ and monoclinic $(b)$ structure of $Zr_{(1-2x)}Y_{2x}O_{(2-x)}V_x$,
          most stable configuration at $x=0.03125$. The forces acting on the ions in the ``starting geometry'',
          i.e. with the atoms placed at the coordinates of the relaxed $ZrO_2$ system, are represented with blue arrows. 
          In the tetragonal structure the oxygen atoms ($O$, small red spheres) closer to the oxygen vacancy
          ($V$ represented as a small black sphere in the figure) move toward it of \ $\approx 0.3 - 0.5$ \AA, while
          the Zirconium atoms ($Zr$, big dark/magenta sphere) next nearest neighbor move away from it of \ $\approx 0.1 - 0.2$ \AA\
          from the ``starting geometry''.
          The Yttrium atom ($Y$, biggest light/yellow sphere) is almost fix. The behavior is similar in the monoclinic structure, though
          the vacancy has just 3 nearest neighbor here, one of which is an Yttrium atom.
          The Yttrium atom moves away from the vacancy suggesting that a charged--defects interaction model 
          based on electrostatic is not correct and that exchange--correlation effects play a major role.}
 \label{fig:structures}
\end{figure}

The most favored configuration, for both the (T) and the (M) phase, is represented in Fig.~\ref{fig:structures},
together with the forces acting on each atom in the initial geometry of pure 
$ZrO_2$.
The $Y$ ions are not attracted by the oxygen vacancies and indeed in the (M) phase they are pushed
away from the vacancy. Hence both the analysis of the energetically favored configuration and the internal relaxation
of the ions does not follow the intuitive scheme of charged defects with the
vacancies a $+2e$ and the $Y$ ion a $-1$ charged sites, suggesting that exchange--correlation effects play a
role in the stabilization mechanism.

\begin{figure}[t]
 \begin{center}
 \subfigure[Starting geometry]{\includegraphics[width=0.48\textwidth]{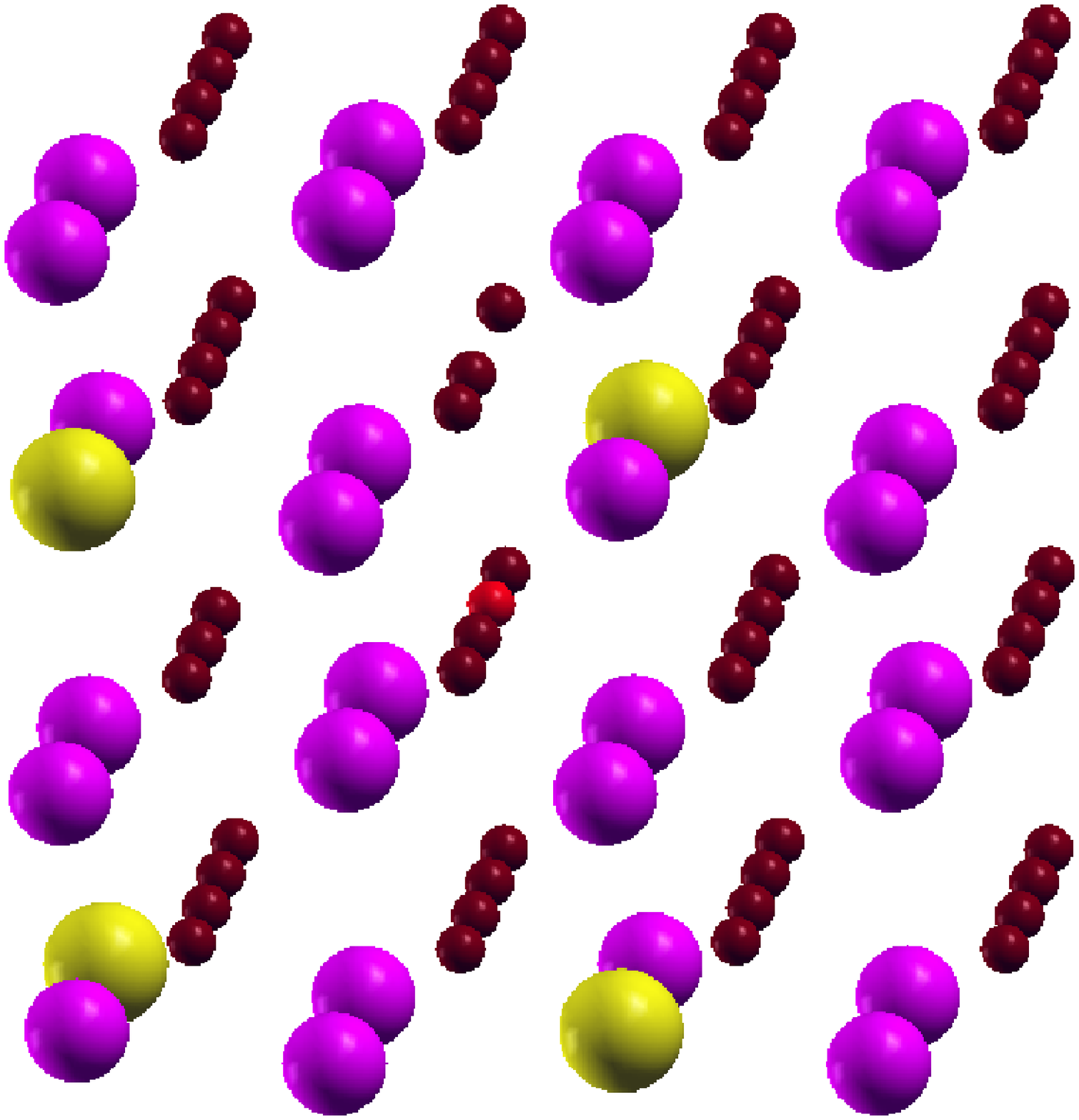}\label{fig:Oxygen_moves_00}}
 \subfigure[Relaxed geometry]{\includegraphics[width=0.48\textwidth]{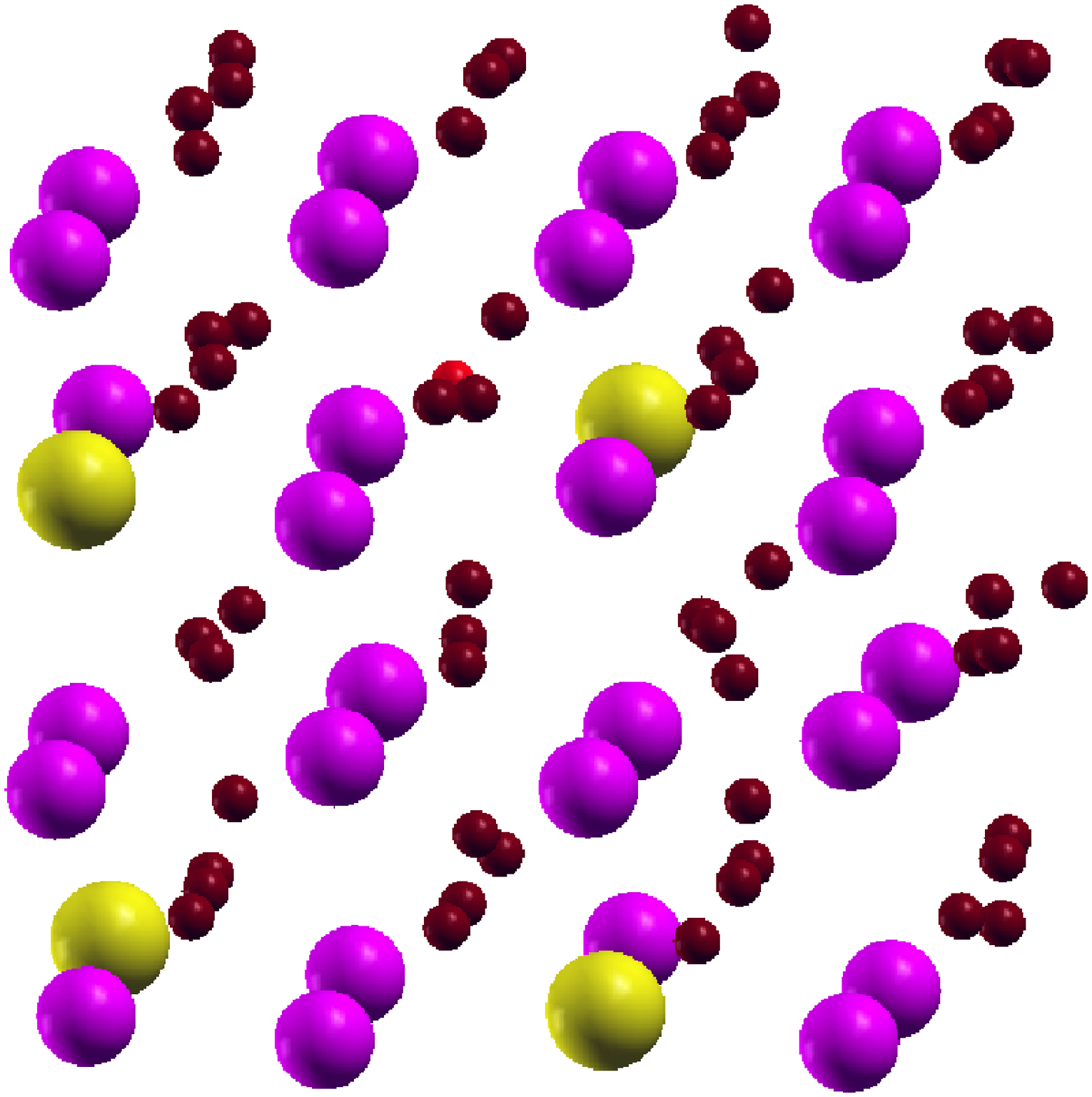}\label{fig:Oxygen_moves_last}}
 \end{center}
  \linespread{1.}
 \caption{(colors online) Starting ($a$) and final relaxed geometry ($b$) of the $Zr_{1-x}O_{2-x/2}Y_{x}V_{x/2}$ system with $x=12.5\%$.
          $O$ atoms are dark (red) small spheres, $Zr$ atoms are big dark spheres (magenta), and $Y$ atoms are biggest light spheres
          (Yellow). The oxygen which changes position is a small light (light red) sphere. Before relaxation the two vacancies are
          aligned along the $[111]$ direction at a distance of $8.45$ Bohr; they can be detected as in the figure the oxygens
          are aligned in groups of four atoms everywhere except where a vacancy exist.
          After the relaxation two vacancies are aligned along the $[101]$ direction at a distance of $\approx 6.2$
          Bohr (this is the distance that the missing oxygens would have at the geometry of pure $ZrO_2$).}
 \label{fig:vacancy_moves}
\end{figure}

Also the relative vacancy--vacancy orientation and position is in contrast with an electrostatic based model.
In the (C)/(T) phase, while Stapper et al.\cite{Stapper1999} based their calculation on the assumption that oxygen vacancies
should remain as far as possible, Ostanin et al.\cite{Ostanin2002A} results suggest that oxygen vacancies tend to couple
along the $[111]$ direction with a single $Zr$ ion in between, thus at a relative distance of about
$\approx 4.47$ \AA. Also experimental X--ray data~\cite{Goff1999} seems to support this idea. No data 
about the orientation of oxygen vacancies in the (M) phase is present in the literature.

Changing the relative position of di--vacancy complex in our model (i.e. two vacancy at $x=12.50\%$)
we found differences in total energy of $\Delta E_{tot}\approx 1.1$ eV (per di--vacancy).
The oscillations in the mean field (MF), i.e. the sum of the kinetic~\cite{NoteA} and electrostatic energy,
and the exchange--correlation (xc) component of the energy are much more pronounced, with
$\Delta E_{xc}\approx 8.2$ eV and $\Delta E_{MF}\approx 7.2$ eV. We will discuss this more in detail in the next section.
Our results are, in some aspects, different from the ones suggested by the two cited works.
Indeed we found that, while vacancies does not repel, the configuration suggested as most favorable for the (T) structure by
Ostanin et al.\cite{Ostanin2002A}, is not the lowest in energy. Indeed, surprisingly, the latter comes out to be unstable,
with one of the two vacancies which changes position if we let the
system relax (Fig.~\ref{fig:vacancy_moves}). This last result suggests that, in presence of vacancies, oxygen atoms can
move with (nearly) no potential barrier to overcome. This is a theoretical evidence that YSZ is a good ionic 
conductor. Indeed this property is important for many applications and has been
investigated in other works~\cite{Pietrucci2008}.

We also considered some cases at atomic doping concentrations of $x=18.75\%$
and one test case at $x=25.00\%$. Both for $x=12.50\%$ and $x=18.75\%$ there is a huge number of possible relative
position of the oxygens and we did not try to systematically explore which is the best configuration.
At $x=12.25\%$ the best found configuration has two vacancies aligned in the $xy$ plane.
A simple explanation of this fact could be provided considering the axial anisotropy of the dielectric
constant ($\kappa$) of the (T) phase. The component $\kappa_{xx}=\kappa_{yy}$ are larger than the $\kappa_{zz}$,
thus providing a more effective screening of charged oxygen vacancies placed 
along the $xy$ plane. However we did not find this to be the sole mechanism, as some of the other configurations
with the vacancies aligned in the $xy$ plane have higher energy than some configurations with the vacancies aligned, for example,
in the $xz$ plane.

To conclude this section, the behavior of the oxygen vacancies present some differences, and some similarities, between the
two phases of Zirconia. A key important difference however is that the formation of vacancies is more favored in the (T) phase
than in the (M) phase as one can see from Fig.~\ref{fig:vac_energy}. This fact indeed implies that $Y_2O_3$ doping tends 
to stabilize the (T) phase.

\subsection{Theoretical stabilization: $xc$--energy and vacancies.}

\begin{figure}[t]
 \begin{center} 
   \includegraphics[width=0.95\textwidth]{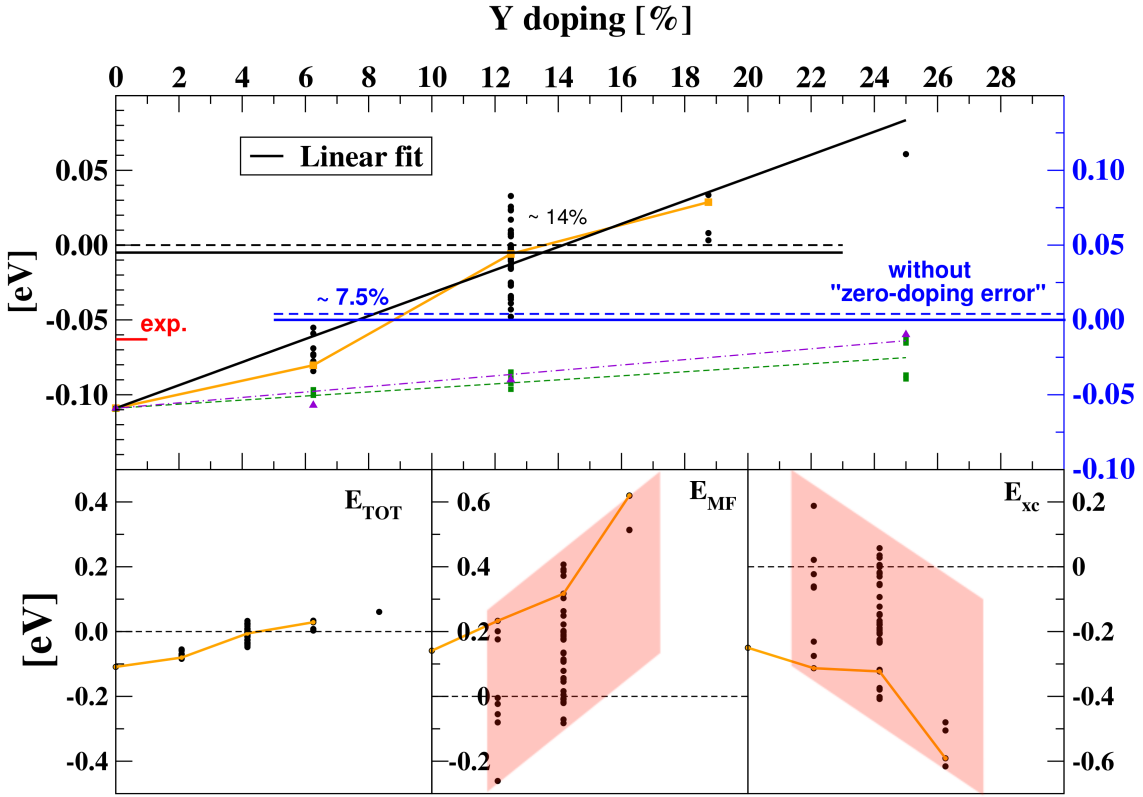}
 \end{center}
  \linespread{1.}
 \caption{(colors online) The difference in total energy per molecular unit (up panel) between the monoclinic and
          the tetragonal phase decrease increasing the $Y_2O_3$ doping (black circles). The result is not simply
          the sum of the effect induced by $Y$ doping without vacancies (violet triangles, dot dashed line
          for the linear fit) and oxygen vacancies without Yttrium (green squares, dashed line for the linear fit).
          When no Yttrium is present the value of the $x$ coordinate is chosen so that the number of vacancies
          for the $Y_2O_3$ doping case is equal to the number of vacancies for the oxygen vacancies only case.
          In the bottom panels the variations in the total energy as a function of the configurations is compared with
          the variations in the mean field and exchange--correlation energy. The pink shadowed area are a guide
          for the eyes, to highlight the trend of the two energy components.}
 \label{fig:E_total}
\end{figure}

The configuration explored allow to model the (M)$\rightarrow$(T) DIPT
as a function of the Yttria doping. In Fig.~\ref{fig:E_total} we see that the energy difference between the (T)
and the (M) phase decrease increasing the doping concentration. This is a clear signature that we are correctly describing
the DIPT. However the stabilization of the (T) phase
happens, after a linear fit of our data, at higher doping concentration, $\approx 14$\%, than the experimental value
$\approx 7$\%~\cite{Lamas2000}.
The main error is due to an overestimation of the zero--doping energy difference of $0.056$ eV/m.u.
($E_{exp}\approx 0.063$ eV/m.u.~\cite{Achermann1975} while $E_{DFT}\approx0.109$ eV/mu.u.). Indeed it is reasonable to assume
that the trend of the energy difference is better computed than its absolute value and accordingly assuming a
constant ``zero--doping error'' for every $Y$ concentration we can subtract it (the zero energy in Fig.~\ref{fig:E_total} shifts from
the black line to the blue line) obtaining $x_{DIPT}\approx 7.5\%$,
which is in excellent agreement with the experimental value. Similar results are obtained if one consider, instead
of the linear fit, the energy difference between the two best configurations (orange dashed line in Fig.~\ref{fig:E_total}).

\begin{figure}[t]
 \begin{center}
   \includegraphics[width=0.99\textwidth]{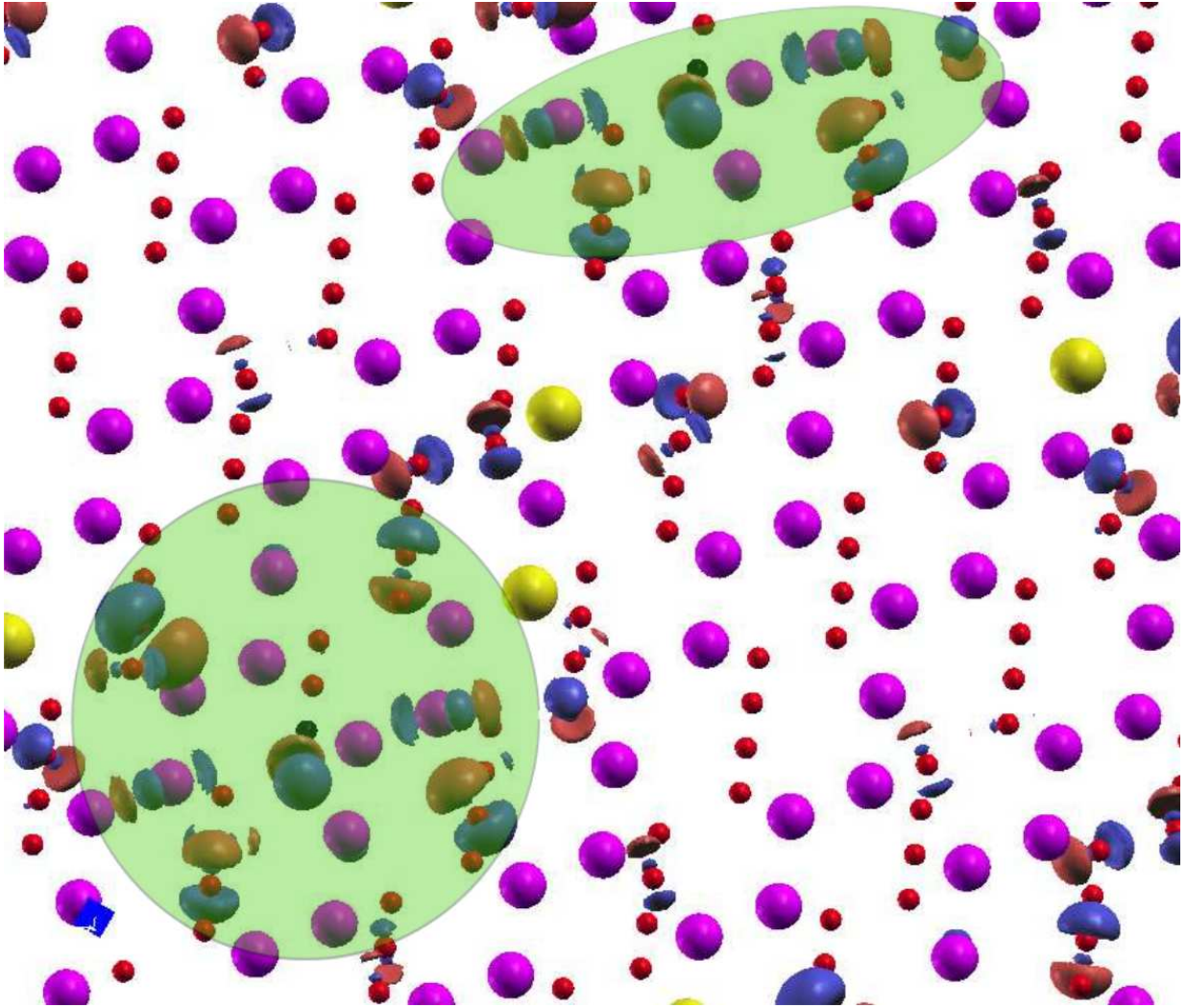}
 \end{center}
  \linespread{1.}
 \caption{(colors online) 3D plot of the polarization charge, in the tetragonal phase, defined as:
          $\rho_{pol}=[32\rho(Zr_{0.94}O_{1.97}Y_{0.06}V_{0.03})+\rho(O)]-32\rho(ZrO_2)$:
          iso--surfaces for $\rho_{pol}=\pm\rho_0$ are represented,
          with the positive ones in red (lighter), the negatives in blue (darker). Oxygen atoms are the small spheres (red),
          Zirconium atoms the dark big spheres (magenta) and Yttrium atoms the light biggest spheres (yellow).
          The (green) shadowed regions highlight the regions around an oxygen vacancy (small black sphere in 
          center of the region), where $\rho_{pol}$ is different from zero.}
 \label{fig:rho_pol}
\end{figure}

For the pure $ZrO_2$ we have also computed the contribution to the Helmholtz free energy due to thermal excitations
of phonons. We found that the difference in the thermal contribution is negligible at room temperature, while the
difference of the zero point phonon energy between the two phases is $\approx 0.005$ eV/m.u.\ , with a small
effect on the value of $x_{DIPT}$ (In Fig.~\ref{fig:E_total} this amounts to the shift of the zero from the dashed
to the continuous line).

It is interesting to compare how the mean field energy,
$E_{MF}[\rho]=T[\rho]+V^{ext}+V_H[\rho]$, and the exchange--correlation energy, $E_{xc}[\rho]$, change
considering different configurations. 
In Fig.~\ref{fig:E_total}, bottom panels, we see that the two component oscillate
about 10 times more than the total energy as a function of the chosen atomic configuration.
The mean field energy, in pure $ZrO_2$, is lower in the (T) phase
(Fig.~\ref{fig:E_total}, central bottom panel); 
that is, if the xc--energy is neglected, the (T) phase is thermodynamically the most 
stable. The inclusion of $Y$ doping and oxygen vacancies enhance this aspect as,
on average $d (\Delta E_{MF})/d x> 0$, with $x$ the atomic doping content and $\Delta E_{MF}=E_{MF}^{(M)}-E_{MF}^{(T)}$.
This is likely because the tetragonal phase, which has an higher dielectric constant, better
screens the electrostatic perturbation $V^{ext}_{doping}$ induced by the doping~\cite{NoteE}.

In order to have a picture of the screening effect we have defined (see Fig.~\ref{fig:rho_pol} and
caption) the polarization charge, that is the difference in the electronic density between the pure
and the doped system. In Fig.~\ref{fig:rho_pol} we see that the main perturbation is induced by
the oxygen vacancies. The screening mechanism however is very efficient and accordingly
the value of the polarization charge is sensibly different from zero only
on the atoms NN to the vacancies.

The $xc$--energy in pure $ZrO_2$ instead is lower in the (M) phase and 
on average $d \Delta E_{xc}/d x< 0$.
To understand this point we consider 
the perturbations induced in the electronic part of the Hamiltonian of
pure $ZrO_2$, that is $V^{pert}=V^{ext}_{doping}+V^{ind}_{ions}$, where $V^{ind}_{ions}$ 
is the additional perturbation generated from the dislocation of the atoms.
$V^{pert}$ has the shape of a random potential in a perfect periodic system which
tends to destroy the collective behavior of the electron gas and thus electrons lose correlation.
This is not an ``on site'' correlation, but rather a ``ranged'' correlation energy
and so we expect that GGA can better describe this effect than LDA.
While $V^{ext}_{doping}$ is similar for the two phases, i.e. we are considering the
same kind of doping,
the $V^{ind}_{ions}$ term is greater in the (T) phase due to its higher screening
(see note \onlinecite{NoteE}). Accordingly the $xc$--energy is more
penalized, that is $d E^{(M)}_{xc}/d x< d E^{(T)}_{xc}/d x$. This explains $d \Delta E_{xc}/d x< 0$
and thus we can infer that in the screening process electrons lose $xc$--energy.

The emerging picture is that the (M) $\rightarrow$ (T) DIPT is a balance
between different mechanisms with a key role played by vacancies~\cite{NoteB}, though the
role of Yttrium atoms is not only to induce vacancies (see Fig.~\ref{fig:E_total}
and caption).

\section{Conclusions}
The phase transition of Zirconia induced with Yttria doping is very well known experimentally and indeed
Yttria stabilized Zirconia is commonly used in many applications. Our results show that an ab--initio approach
is able to correctly describe this mechanism, with a predicted phase transition at a doping concentration
of $\approx 7.5\%$, in good agreement with experimental data. This result confirms the opportunity to predict
the effect of other less characterize kind of doping which could be of potential interest for new
applications, as for example doping with magnetic materials~\cite{Ostanin2007}.

Moreover, within density functional theory we can explore how
physical properties of the system are influenced by doping.
As an example we considered how different components of the energy changes and in particular the
behavior of the exchange--correlation energy. We showed that oxygen defects play a major role in
the phase transition and how the perturbation induced by oxygen vacancies to the system is screened
in the high-k tetragonal phase. The same approach could be used to check the effect of doping on
other quantities, which are not easily accessible experimentally; among others,
the value of the dielectric constant, which has a key importance for applications in
the field of micro--electronics.

Finally these results, improving our understanding of the phase transition, can also be used to tune
the parameters of models for the description of realistic devices based on Zirconia, whose dimensions
are still beyond the capabilities of first--principles simulations.

\section{Acknowledgments}
This work was funded by the Cariplo Foundation through the OSEA project (n.~2009-2552).
The authors would like to acknowledge
Professor Giovanni Onida and the ETSF~\cite{ETSF} Milan node for the opportunity of running
simulations on the ``etsfmi cluster''. We also acknowledge computational resources provided
under the project OSEA by the Consorzio Interuniversitario per le Applicazioni di
Supercalcolo Per Universit\'a e Ricerca (CASPUR).



\begin{thebibliography}{99}
\bibitem{NoteD} Also for the doped case we considered only the (M) and the (T) phases. Indeed the (C) phase is a special
(T) phase with $c/a=1$ and $d_z=0$. The description of this phase is included in our work as we let the atoms relax their
position and, in few cases, also the cell parameters.
\bibitem{NoteC} Thought these are not the physical DOS, in other works~\cite{Jiang2010,Gruning2010} it has been shown that
for $ZrO_2$ the main difference with the real DOS amounts in a red shift of the gap as common within DFT/LDA.
\bibitem{NoteA} Here what we consider is the KS kinetic energy.
\bibitem{NoteE} Here we are studying the response of pure $ZrO_2$ to a static
external perturbation $V^{ext}_{doping}$.
The self--consistent solution, in the Bohr--Oppenheimer approximation, of the ionic plus the electronic
Hamiltonian $H^{I+el}_{ZrO_2}[\rho]+V^{ext}_{doping}$ describes the full response of the
system to this perturbation. For a comparison the static
dielectric $\epsilon(0)$ constant of pure $ZrO_2$ can give an insight on how the system will react to this
perturbation to linear order. In the self--consistent solution the phononic contribution to the dielectric constant, $\epsilon_{ph}(0)$,
is accounted by the relaxation of the atomic positions, while the electronic contribution, $\epsilon_{el}(0)$, is acconted
by the the solution electronic part of the Hamiltonian $H^{el}_{ZrO_2}[\rho]+V^{ext}_{doping}+V^{ind}_{ions}$.
In high--k materials like Zirconia the contribution from the lattice is dominant.
The (T) phase has higher dielectric constant and accordingly we expect a greater distortion
of the crystal, provided that the linear order behavior is respected to all orders.
\end{thebibliography}
\end{document}